\documentclass[preprint,aps]{revtex4}

\usepackage{graphicx}
\usepackage{dcolumn}
\usepackage{bm}
\usepackage{epsf}

\newcommand{\eq}[1]{eq.~(\ref{#1})}
\newcommand{\eqs}[2]{eqs.(\ref{#1},\ref{#2})}

\newcommand{\Eq}[1]{Eq.~(\ref{#1})}

\newcommand{\ur}[1]{(\ref{#1})}
\newcommand{\urs}[2]{(\ref{#1},\ref{#2})}

\newcommand{\beq}{\begin{equation}}
\newcommand{\eeq}{\end{equation}}
\newcommand{\ee}{\epsilon}
\newcommand{\la}[1]{\label{#1}}
\newcommand{\bea}{\begin{eqnarray}}
\newcommand{\eea}{\end{eqnarray}}
\newcommand{\beqa}{\begin{eqnarray}}
\newcommand{\eeqa}{\end{eqnarray}}
\newcommand{\ba}{\begin{array}}
\newcommand{\ea}{\end{array}}

\newcommand{\half}{{\textstyle{\frac{1}{2}}}}

\newcommand{\at}{\overline{10}}

\newcommand{\nn}{\nonumber}

\renewcommand{\d}{\dagger}

\newcommand{\scp}[2]{{\bf #1}\cdot{\bf #2}}
\renewcommand{\a}[2]{a_{#1}({\bf #2})}

\renewcommand{\u}[2]{u_{#1}({\bf #2})}

\renewcommand{\v}[2]{v_{#1}({\bf #2})}
\newcommand{\vd}[2]{\bar v^{#1}({\bf #2})}
\newcommand{\inte}[1]{\int\!(d{\bf #1})}

\def\Dirac#1{#1\hskip-5pt/}
\def\dd{\Dirac\partial}

\renewcommand{\slash}[1]{\Dirac #1}

\begin{document}

\title{\bf Baryons as Fock states of 3,5,... quarks}

\author{
\bf Dmitri Diakonov$^{1,2,3}$ and Victor Petrov$^3$}
\affiliation{
$^1$ Thomas Jefferson National Accelerator Facility, Newport News, VA 23606, USA\\
$^2$ NORDITA, Blegdamsvej 17, DK-2100 Copenhagen, Denmark \\
$^3$ St. Petersburg Nuclear Physics Institute, Gatchina, 188 300, St. Petersburg, Russia
}

\date{September 29, 2004}

\begin{abstract}
We present a generating functional producing quark wave functions of all 
Fock states in the octet, decuplet and antidecuplet baryons in the mean field 
approximation, both in the rest and infinite momentum frames. In particular, 
for the usual octet and decuplet baryons we get the $SU(6)$-symmetric wave functions 
for their 3-quark component but with specific corrections from relativism and 
from additional quark-antiquark pairs. For the exotic antidecuplet baryons 
we obtain the 5-quark wave function~\cite{footnote1}.
\end{abstract}

\pacs{12.38.-t, 12.39.-x, 12.39.Dc, 14.20-c} 
\keywords{baryons, chiral symmetry, Fock states, exotic baryons}

\maketitle

\section{Introduction}
\label{sect1}
It is a great pleasure for us to write a paper in honor of our old friend
Klaus Goeke who has made an enormous contribution to the development of the
Chiral Quark Soliton Model (CQSM). 

We have noticed from experience that many people hearing the word ``soliton''
immediately imagine the XIX century English gentleman racing after a solitary
wave along the Thames, and are for that reason forever scared off from the model.
In more recent times, another English gentleman T.H.R.~Skyrme suggested that
nucleons can be viewed as solitons of the pion field. This scares some people even 
more as few would imagine a fermion built from a bosonic field. This paper
is another attempt to persuade those people that the CQSM is about {\bf quarks}:
we are going to present explicit quark wave functions following in particular
from that model. Few people call atoms or nuclei solitons; however, they {\it can be}
called solitons of the electrostatic and of the mean nuclear field, respectively.
Maybe we should rename the CQSM into the Mean Field Approximation, to make
it sound more traditional and less challenging the common sense.    

The mean field approach to bound states is usually justified by the large
number of participants. The Thomas--Fermi approximation to atoms is justified 
at large $Z$, and the shell model for nuclei is justified at large $A$. 
In baryons, the appropriate large parameter justifying the mean field approach
would be the number of colors $N_c$~\cite{Witten}. There are two kind of corrections in
$1/N_c$. One kind is due to the fluctuations of the chiral field about its
mean-field value in a baryon. These are loop corrections and are additionally
suppressed by factors of $1/(2\pi)$. With the present precision, such corrections
can be ignored. The second type can be called kinematical: they are due to the
rotations of a baryon, and are not suppressed additionally. Many baryon observables 
get such corrections, the isovector magnetic moment or the axial constant being 
examples~\cite{corrNc1,corrNc2}. For example, $g_A$ of the nucleon gets a correction 
factor $(1+2/N_c)$ equal to 5/3 in the real world. Such kind of corrections should 
be collected, if possible. When and if this is done, the CQSM or else the Mean Field 
Approximation for baryons becomes quite precise. 

It should be noted that these anomalously large corrections arise from the imaginary  
part of the effective chiral action~\cite{DP-CQSM}, and are absent {\it e.g.} 
in the Skyrme model. Being important in ordinary baryons, the anomalously large $1/N_c$
corrections become crucial for exotic baryons. In the Skyrme model the exotic
$(\at,\half^+)$ baryon antidecuplet does not exist unless one extends 
the parameters of the model~\cite{KlebanovRho}; even if this is done, the exotic
$\Theta^+$ baryon appears as a broad resonance. Meanwhile in the CQSM, the formally
leading term for the $\Theta$ width is strongly cancelled by terms non-existing in 
the Skyrme model~\cite{DPP97}, and this cancellation pertains at any $N_c$~\cite{Prasz-Theta-Nc}.  

Therefore, it would be helpful to have a formalism which relies on the mean chiral
field but treats the rotational corrections exactly at any $N_c$ and at the real-world
$N_c\!=\!3$ in particular. In fact, this was the logic always adopted in Bochum.
Its philosophy has been recently reviewed in Ref.~\cite{D04}. We shall apply it here 
to reveal the 3,5,7...-quark wave functions of the octet, decuplet and antidecuplet baryons.

A schematic view of baryons in the Mean Field Approximation is presented in Fig.~1, 
and it is within this construction all observables have been so far computed in the CQSM. 

\begin{figure}[htb]
\begin{minipage}[t]{.50\textwidth}
\includegraphics[width=\textwidth]{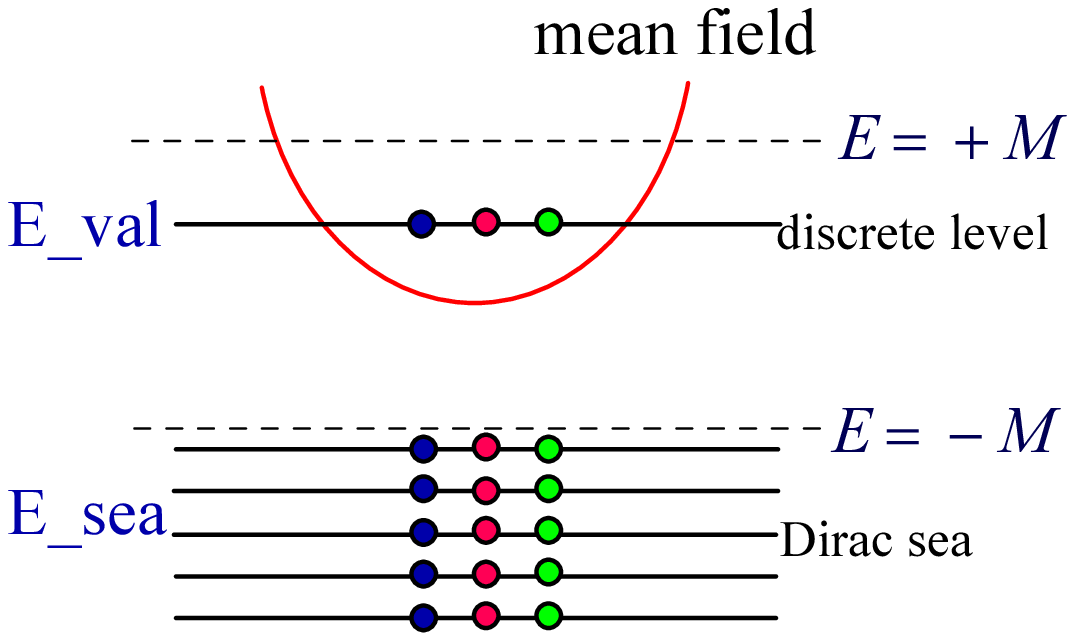}
\caption{A schematic view of baryons in the Mean Field Approximation. There are three ``valence''
quarks at a discrete energy level created by the mean field, and the negative-energy Dirac
continuum distorted by the mean field, as compared to the free one.}
\label{fig:2}
\end{minipage}
\hfil
\begin{minipage}[t]{.45\textwidth}
\includegraphics[width=\textwidth]{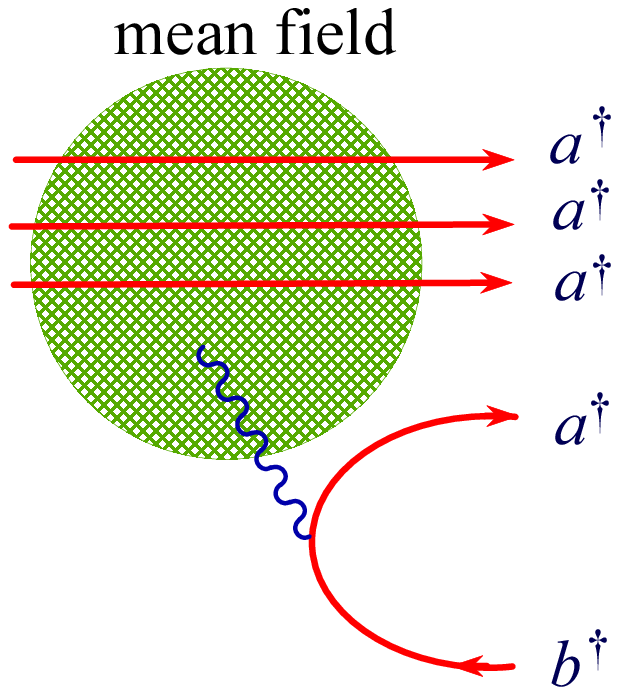}
\caption{Equivalent view of baryons in the same approximation, where the distorted Dirac sea
is presented as quark-antiquark pairs. Their wave function is given by the quark Green function
in the background mean field, at equal times.}
\label{fig:3}
\end{minipage}
\end{figure}

An alternative but mathematically equivalent description has been recently suggested by
one of the authors and Polyakov~\cite{PP-IMF}. If the mean field makes the Dirac sea
less dense than in the vacuum at certain momenta, it means a hole or the presence of an 
{\em antiquark with positive energy}. If the Dirac sea is more dense in some partial 
wave and at some momenta, it means the presence of an additional {\em negative-energy quark} 
with the corresponding quantum numbers. Since the total number of levels in the sea 
is its baryon number and is conserved whatever the background field, it implies that 
any distortion of the Dirac sea by the mean field creates an equal number of quarks 
and antiquarks or, else, quark-antiquark ($Q\bar Q$) pairs inside a baryon, 
in addition to the three valence quarks, see Fig.~2. Algebraically, the average number 
of $Q\bar Q$ pairs is proportional to the amplitude squared of the mean field, times $N_c$. 
It also depends on the quantum numbers of the baryon in question. For nucleons 
it appears to be somewhat less than one pair on the average. In the antidecuplet, 
the number of $Q\bar Q$ pairs is strictly larger than one. Theoretically, at large $N_c$ 
there are $\sim N_c$ $Q\bar Q$ pairs in any baryon in addition to the $N_c$ valence quarks.

\section{The effective action}

The effective action approximating QCD at low momenta describes ``constituent'' quarks 
with the momentum dependent dynamical mass $M(p)$ interacting with the scalar ($\Sigma$) 
and pseudoscalar (${\bf \Pi}$) fields such that $\Sigma^2+{\bf \Pi}^2=1$ at spatial infinity.
The momentum dependence $M(p)$ serves as a formfactor of the constituent quarks and 
provides the effective theory with the ultraviolet cutoff. Simultaneously, it
makes the theory non-local. The action is~\cite{DP86,DP-CQSM}
\beq
\!S_{\rm eff}\!=\!\!\int\!\frac{d^4pd^4p'}{(2\pi)^8}\,
\bar\psi(p)\left[\slash{p}\,(2\pi)^4\delta^{(4)}(p-p')\!-\!\sqrt{M(p)}
\left(\Sigma(p\!-\!p')+i\Pi(p\!-\!p')\gamma_5\right)\sqrt{M(p')}\right]\psi(p'),
\la{action}\eeq
where $\psi,\bar\psi$ are quark fields carrying color, flavor and Dirac bispinor
indices. In the instanton model of the QCD vacuum from where this action has been
originally derived~\cite{DP86} the function $M(p)$ is such that there is no 
real solution of the mass-shell equation $p^2=M(-p^2)$, therefore quarks are 
not observable as asymptotic states, -- only their bound states. However, this is not the true 
confinement. Unfortunately, the instanton model's $M(p)$ has a cut at $p^2=0$ 
corresponding to massless gluons left in that model. In the true confining theory 
there should be no such cuts. Nevertheless, such $M(p)$ creates some kind of a
soft ``bag'' for quarks. Contrary to the primitive bag picture which violates
all principles one can think of, \eq{action} supports all general principles, like
relativistic invariance and sum rules for conserved quantities. 
  

Turning to baryons, the mean $\Sigma,\Pi$ field (called chiral field for short in what 
follows) in the full non-local formulation \ur{action} has been found by Broniowski,
Golli and Ripka~\cite{BGR}. It sets an example how one has to proceed in the model
calculations. However, to simplify the mathematics we shall use here a more standard
approach: we shall replace the constituent quark mass by a constant and mimic the
decreasing function $M(p)$ by the UV Pauli--Villars cutoff~\cite{SF}. 
 
\section{Baryon wave function in terms of quark creation-annihilation operators}

Let $a,a^\dagger({\bf p})$ and $b,b^\dagger({\bf p})$ be the annihilation--creation
operators of quarks and antiquarks (respectively) of mass $M$, satisfying the usual 
anticommutator algebra $\{a({\bf p})a^\d({\bf p'})\}=\{b({\bf p})b^\d({\bf p'})\}
=(2\pi)^3\delta^{(3)}({\bf p}-{\bf p'})$ with $a,b|0\!\!>=\!0,\,<\!\!0|a^\d,b^\d\!=\!0$. 
For quarks, the annihilation-creation operators carry, apart from the 3-momentum ${\bf p}$,
also the color $\alpha$, flavor $f$ and spin $\sigma$ indices but we shall
suppress them until they are explicitly needed. The Dirac sea is presented by 
the coherent exponent of the quark and antiquark creation operators,
\beq
{\rm coherent\;exponent}=\exp\left(\inte{p}(d{\bf p'})\,a^\dagger({\bf p})\,
W({\bf p},{\bf p'})\,b^\dagger({\bf p'})\right)|0\!>, 
\la{cohexp}\eeq   
where $(d{\bf p})=d^3{\bf p}/(2\pi)^3$ and $W({\bf p_1},{\bf p_2})$
is the quark Green function at equal times in the background $\Sigma,{\bf \Pi}$ 
fields, see Fig.~2. In the saddle point approximation these fields
are replaced by the mean field:
\beq
\pi({\bf x})={\bf n\cdot\tau}P(r),\qquad {\bf n}={\bf x}/r,\qquad \Sigma({\bf x})=\Sigma(r).
\la{hh}\eeq 
On the chiral circle $\Pi={\bf n\cdot\tau}\,\sin P(r),\;\Sigma(r)=\cos P(r)$ where
$P(r)$ is the profile function of the self-consistent field. 
We shall specify the pair wave function $W({\bf p},{\bf p'})$ below. 

It is assumed that the self-consistent chiral field creates a bound-state level 
for quarks, whose wave function $\psi_{\rm lev}$ satisfies the static Dirac equation
with eigenenergy $E_{\rm lev}$~\cite{KR,BB,DP-CQSM}:
\beq
\psi_{\rm lev}({\bf x})=\left(\ba{c}\ee^{ji}h(r)\\-i\ee^{jk}\,({\bf \sigma\!\cdot\!n})^i_k\,j(r)
\ea\right),\qquad\left\{\ba{c}h'+h\,M\sin P-j(M\cos P+E_{\rm lev})=0,\\
j'+2j/r-j\,M\sin P-h(M\cos P-E_{\rm lev})=0.\ea\right.
\la{level}\eeq
In the non-relativistic limit ($E_{\rm lev}\approx M$) the $L\!=\!0$ upper component 
of the Dirac bispinor $h(r)$ is large while the $L\!=\!1$ lower component $j(r)$ is small.

The valence quark part of the baryon wave function is given by the product 
of $N_c$ quark creation operators that fill in the discrete level:
\beqa\la{val1}
&&{\rm valence}=\prod_{{\rm color}=1}^{N_c}\int(d{\bf p})\,F({\bf p})\,a^\dagger({\bf p}),\\    
&&F({\bf p})\!=\!\!\int\!(d{\bf p'})\sqrt{\frac{M}{\ee_p}}\!\left[\bar u({\bf p})
\,\gamma_0\,\psi_{\rm lev}({\bf p})\,(2\pi)^3\delta({\bf p}\!
-\!{\bf p'})\!-\!W({\bf p},{\bf p'})\,\bar v({\bf p'})\,\gamma_0\,\psi_{\rm lev}(-\!{\bf p'})\right]\!,
\la{val2}\eeqa
where $\psi_{\rm lev}({\bf p})$ is the Fourier transform of \eq{level}.
The second term in \Eq{val2} is 
the contribution of the distorted Dirac sea to the one-quark wave function. 
$u_\sigma({\bf p})$ and $v_\sigma({\bf p})$ are 
the plane-wave Dirac bispinors projecting to the positive and negative frequencies, 
respectively. In the standard basis they have the form
\beq\la{uv}
\u{\sigma}{p}=\left(\ba{c}\sqrt{\frac{\ee+M}{2M}}s_\sigma\\
\sqrt{\frac{\ee-M}{2M}}\frac{{\bf p\cdot\sigma}}{|p|}s_\sigma\ea\right),\qquad
\v{\sigma}{p}=\left(\ba{c}\sqrt{\frac{\ee-M}{2M}}\frac{{\bf p\cdot\sigma}}{|p|}s_\sigma\\
\sqrt{\frac{\ee+M}{2M}}s_\sigma\ea\right),\qquad \bar u u=1=-\bar v v\,,
\eeq
where $\ee=\!+\!\sqrt{{\bf p}^2+M^2}$ and $s_\sigma$ are two 2-component spinors 
normalized to unity, for example,
\beq
s_1=\left(\ba{c}1\\0\ea\right),\qquad s_2=\left(\ba{c}0\\1\ea\right),\qquad \sigma=1,2.
\la{s}\eeq

The full baryon wave function is given by the product of the valence part \ur{val1} 
and the coherent exponent \ur{cohexp} describing the distorted Dirac sea. 
Symbolically, one writes the baryon wave function in terms of the quark and 
antiquark creation operators ~\cite{PP-IMF}:
\beq
B[a^\d,b^\d]=\prod_{{\rm color}=1}^{N_c}\int(d{\bf p})\,F({\bf p})\,a^\dagger({\bf p})\;
\exp\left(\inte{p}(d{\bf p'})\,a^\dagger({\bf p})\;
W({\bf p},{\bf p'})\,b^\dagger({\bf p'})\right)|0\!>.
\la{B1}\eeq 

At this point one has to recall that the saddle point at the self-consistent chiral 
field is degenerate in global translations and global $SU(3)$ flavor rotations [the
$SU(3)$ breaking by the strange mass can be treated as a perturbation.] Integrating
over translations leads to the momentum conservation: the sum of all quarks and antiquarks
momenta have to be equal to the baryon momentum. Integration over rotations $R$ leads
to the projection of the flavor state of all quarks and antiquarks onto the spin-flavor
state $B(R)$ describing a particular baryon from the $\left({\bf 8},\frac{1}{2}^+\right), 
\left({\bf 10},\frac{3}{2}^+\right)$ or $\left({\bf \overline{10}},\frac{1}{2}^+\right)$
multiplet. 

Restoring color ($\alpha=1,2,3$), flavor ($f=1,2,3$), isospin ($j=1,2$) and spin 
($\sigma=1,2$) indices, the quark wave function inside a particular baryon $B$ with spin 
projection $k$ is given, in full glory, by 
\beqa\nn
&&\Psi^B_k=\int \!dR\,B^*_k(R)\,\ee^{\alpha_1\alpha_2\alpha_3}\prod_{n=1}^{3}
\int\!(d{\bf p_n})\,R^{f_n}_{j_n}\,F^{j_n\sigma_n}({\bf p_n})\,
a^\dagger_{\alpha_nf_n\sigma_n}({\bf p_n})\\
\la{Psi}
&&\cdot\exp\left(\int\!(d{\bf p})(d{\bf p'})\,a^\dagger_{\alpha f\sigma}({\bf p})\,R^f_j\,
W^{j\sigma}_{j'\sigma'}({\bf p},{\bf p'})\,R^{\dagger\,j'}_{f'}\,
b^{\dagger\,\alpha f'\sigma'}({\bf p'})\right)|0\!\!>\,.
\eeqa  
This is the ``generating functional" mentioned in the Abstract. Expanding the coherent
exponent to the 0$^{\rm th}$, 1$^{\rm st}$, 2$^{\rm nd}$... order one reads off the
3-, 5-, 7-... quark wave functions of a particular baryon from the octet, decuplet or
antidecuplet. 

To make this powerful formula fully workable, we need to give explicit expressions
for the baryon rotational states $B(R)$, the valence wave function $F^{j\sigma}({\bf p})$
and the $Q\bar Q$ wave function in a baryon $W^{j\sigma}_{j'\sigma'}({\bf p},{\bf p'})$. 

\subsection{Baryon rotational states}   

In general, baryon rotational states $B(R)$ are given by the $SU(3)$ Wigner 
finite-rotation matrices~\cite{hyperons}, and any particular projection 
can be obtained by a routine $SU(3)$ Clebsch--Gordan technique. However, 
in order to see the symmetries of the quark wave functions it is helpful to use
explicit expressions for $B(R)$, and integrate over the Haar measure in \eq{Psi}
explicitly. 
   
Let us give a few examples of the baryons' (conjugate) rotational wave functions $B^*(R)$:
\beqa
\la{p} 
&&{\rm proton,\;spin\;projection}\;k:\qquad p^*_k(R)=\sqrt{8}\,\ee_{kl}R^{\dagger\,l}_1R^3_3,\\
\la{n} 
&&{\rm neutron,\;spin\;projection}\;k:\qquad n^*_k(R)=\sqrt{8}\,\ee_{kl}R^{\dagger\,l}_2R^3_3,\\
\la{Deltappuu} 
&&\Delta^{++},\;{\rm spin\;projection}\;+\!
\frac{3}{2}:\qquad \Delta^{++\,*}_{\uparrow\uparrow}(R)=
\sqrt{10}\,R^{\dagger\,2}_1R^{\dagger\,2}_1R^{\dagger\,2}_1,\\
\la{Delta0u} 
&&\Delta^{0},\,{\rm spin\;projection}\,+\!
\frac{1}{2}:\qquad \Delta^{0\,*}_{\uparrow}(R)\!=\!
\sqrt{10}\,R^{\dagger\,2}_2(2R^{\dagger\,2}_1R^{\dagger\,1}_2\!+\!R^{\dagger\,2}_2R^{\dagger\,1}_1),\\
\la{Theta}
&&\Theta^+,\;{\rm spin\;projection}\;k:\qquad \Theta^*_{k}(R)=\sqrt{30}\,R^3_3R^3_3R^3_k,\\
\la{nantiten}
&&{\rm neutron}^*\;{\rm from}\;\at,\;{\rm spin\;projection}\;k:\qquad 
n_{\at,k}^*(R)=\sqrt{10}\,R^3_3(2R^1_3R^3_k+R^3_3R^1_k).
\eeqa
They are normalized in such a way that for any spin projection
\beq
\int\!dR\,B^*_{\rm spin}(R)\,B^{\rm spin}(R)=1.
\la{normB}\eeq
For example, somebody is interested in the quark wave function of the $\Theta^+$ . 
Then one has to substitute \ur{Theta} into the general \eq{Psi} 
and integrate over $R$. If only three valence quarks are taken and the 
coherent exponent is ignored, one gets
\beq
\int\!dR R^3_3R^3_3R^3_k\,R^{f_1}_{j_1}R^{f_2}_{j_2}R^{f_3}_{j_3}=0
\la{no3quarks}\eeq
meaning, of course, that one cannot built the exotic $\Theta^+$ from three quarks. The first
non-zero Fock component of the $\Theta$ is 5Q and is obtained by expanding the $Q\bar Q$ exponent
to the linear order. In this case the appropriate group integral is
\beqa\nn
&&T(\Theta)^{f_1f_2f_3f_4,j_5}_{j_1j_2j_3j_4,f_5,k}
=\int\!dR\,\Theta^*_{k}(R)\,R^{f_1}_{j_1}R^{f_2}_{j_2}R^{f_3}_{j_3}
R^{f_4}_{j_4}R^{\d\,j_5}_{f_5}\\ 
&&=\frac{\sqrt{30}}{180}\,\delta^3_{f_5}\,\delta^{j_5}_k
\left(\ee_{j_1j_2}\ee_{j_3j_4}\ee^{f_1f_2}\ee^{f_3f_4}+
\ee_{j_2j_3}\ee_{j_1j_4}\ee^{f_2f_3}\ee^{f_1f_4}+
\ee_{j_1j_3}\ee_{j_2j_4}\ee^{f_1f_3}\ee^{f_2f_4}\right).
\la{RTheta}\eeqa
It tells us that the antiquark in the $\Theta$ is necessarily $\bar s$ thanks to $\delta^3_{f_5}$,
and that the four quarks are $uudd$ since in the 2-dimensional antisymmetric tensors
$\ee^{f_1f_2}$ and the like, the flavor indices are $f_{1-4}=1,2=u,d$. 

\subsection{The $Q\bar Q$ pair wave function}

As explained in Ref.~\cite{PP-IMF}, the pair wave function 
$W^{j\sigma}_{j'\sigma'}({\bf p},{\bf p'})$ is expressed through the finite-time
quark Green function at equal times in the external chiral field. 
Schematically, it is shown in Fig.~2.
We define the Green function as the solution of the equation
\beq
\left[i\dd-M(\Sigma+i\Pi\gamma_5)\right]_{x_1,t_1}\,G(x_1,t_1|x_2,t_2)=\delta(t_1-t_2)
\delta^{(3)}({\bf x}_1-{\bf x}_2).
\la{G1}\eeq
The quantity $V=M(-1+\Sigma+i\Pi\gamma_5)$ will be called the perturbation 
as due to the non-zero mean field. In what follows we shall rename $\Sigma-1\to\Sigma$.
For the static hedgehog chiral field lying on the chiral circle
\beq
\Sigma^j_{j'}({\bf x})=(\cos P(r)-1)\delta^j_{j'},\qquad 
\Pi^j_{j'}({\bf x})=({\bf n}\cdot\tau)^j_{j'}\,\sin P(r).
\la{SigmaPi}\eeq
We shall need their Fourier transforms,
\beq
\Sigma({\bf q})=\int\!d^3{\bf x}\,e^{-i{\bf q\cdot x}}\,\Sigma({\bf x}),\qquad
\Pi({\bf q})=\int\!d^3{\bf x}\,e^{-i{\bf q\cdot x}}\,\Pi({\bf x}),
\la{SigmaPi_F}\eeq
where $\Sigma({\bf q})$ is real and even while 
$\Pi({\bf q})$ is purely imaginary and odd.
In the frame where a baryon has a constant velocity $v$ along the $z$ axis both fields
get the arguments $(x,y,z)\to \left( x,y,\frac{z-v t}{\sqrt{1-v^2}}\right)$. Both 
fields can be written through the Fourier transforms in the rest frame:
\beq
\Sigma,\Pi\left(x,y,\frac{z-v t}{\sqrt{1-v^2}}\right)=
\inte{q}\,\exp\left(iq_xx+iq_yy+iq_z\frac{z-v t}{\sqrt{1-v^2}}\right)\,\Sigma,\Pi({\bf q}).
\la{SigmaPi_mov}\eeq 

One can present the Green function as a perturbation expansion in $V=\Sigma+i\Pi\gamma_5$:
\beq 
\frac{1}{i\dd-M-V}=\frac{1}{i\dd-M}
+\frac{1}{i\dd-M}V\frac{1}{i\dd-M}+\frac{1}{i\dd-M}V\frac{1}{i\dd-M}V\frac{1}{i\dd-M}+...
\la{Gsymb}\eeq
The important point (used in the derivation of \eq{B1}~\cite{PP-IMF}) is that all 
free Green functions in this equation should be understood with the Feynman $i\ee$ 
prescription in the momentum space, meaning the shift $M-i0$ in the free propagators.

The perturbation \ur{SigmaPi} is in fact very specific: its modulus is always less
than unity. If the pion field is much less than unity, the perturbation is small. 
If the chiral field is not small but has either low ($q\ll M$) or large ($q\gg M$) momenta,
the perturbation is, effectively, also small as will become clear from the final
expression for the pair wave function. Therefore, it is not a bad idea to restrict 
oneself to the first order in the perturbation in $V$, which we are going to do here. 
Keeping higher orders in $V$ has no principle difficulties but in our experience the
first order is usually within 10-15\% from exact (all orders) calculations.     

In the first order in the external field $V$ the Green function is, according to \eq{Gsymb},
\beq
G^{(1)}(x_1,t_1|x_2,t_2)=\int_0^T\!dt\!\int\!d^3z\,
G^{(0)}(x_1,t_1|z,t)\,V(z,t)\,G^{(0)}(z,t|x_2,t_2).
\la{G11}\eeq
Here $T$ is the ``observation time'' during which the external chiral field
exists; it should be put to infinity to obtain the ground-state baryon with given quantum 
numbers. We can write it further in the momentum representation:
\beqa\nn
&&G^{(1)}(x_1,t_1|x_2,t_2)=\int\!(d{\bf q_1})(d{\bf q_1})\,
\exp(-i\scp{q_1}{x_1}-i\scp{q_2}{x_2})\,G^{(1)}({\bf q_1},t_1|{\bf q_2},t_2),\\
\nn
&&G^{(1)}({\bf q_1},t_1|{\bf q_2},t_2)=\int\!\frac{d\omega_1d\omega_2}{(2\pi)^2}\!\int_0^T\!dt'
\!\int\!d^3z\!\int\!(d{\bf q})\,\exp\left[iq_xz_x+iq_yz_y+iq_z\frac{z\!-\!vt'}{\sqrt{1\!-\!v^2}}\right]\\
&&\cdot \exp\left[-i\omega_1(t'\!-\!t_1)+i{\bf q_1}\!\cdot\!{\bf z}
-i\omega_2(t_2\!-\!t')+i{\bf q_2}\!\cdot\!{\bf z}\right]\,
\frac{1}{\slash{q_1}\!-\!M\!+\!i0}V({\bf q})\frac{1}{\slash{q_2}\!-\!M\!+\!i0}
\la{G12}\eeqa
where $q_{1,2\,\mu}=(\omega_{1,2},{\bf q}_{1,2}),\,\slash{q}_{1,2}=\omega_{1,2}\gamma_0
-{\bf q}_{1,2}\cdot{\bf \gamma}$.

The definition of the (conjugate) pair wave function is~\cite{PP-IMF}
\beq
W^{j'\sigma'}_{\!c\,j\sigma}({\bf p'},{\bf p})=-i\sqrt{\frac{\ee\ee'}{M^2}}\,
\left[\vd{\sigma'}{p'}G({\bf p},0|{\bf p'},0)\u{\sigma}{p}\right]
\la{Wc1}\eeq
with the plane-wave bispinors $u,v$ defined in \eq{uv}. One has to integrate
\eq{G12} over $\omega_{1,2}$ and the intermediate point $({\bf z},t')$ where 
the perturbation $V$ acts. Because of the Feynman ``$M-i0$'' rule, one closes 
the integration contour in $\omega_1$ in the lower semiplane and finds the 
contribution of the pole $\omega_1=\ee=\!\sqrt{{\bf p}^2+M^2}$. Integration 
over $\omega_2$ is closed in the upper semiplane with 
$\omega_2=-\ee'=-\!\sqrt{{\bf p'}^2+M^2}$. This is an important although natural 
result: the $Q\bar Q$ pair has a positive-energy antiquark and necessarily 
a negative-energy quark. The physical interpretation, in terms of the level 
density of the Dirac sea, has been given in the Introduction. 

Integration over $d^3z$ leads to the 3-momentum conservation, 
${\bf q}_\perp=-({\bf p}+{\bf p'})_\perp$, $q_z=-(p_z+p'_z)/\sqrt{1-v^2}$. 
Integration over the intermediate time $t'$ gives
the energy denominator $-i/[\ee+\ee'-i0-(p_z+p'_z)v]$. Finally, one has to use the Dirac
equation for the plane-wave bispinors: $(M-\slash{p})\u{\sigma}{p}=0,\;
\vd{\sigma'}{p'}(M+\slash{p'})=0$. As a result one obtains
\beq
W^{j'\sigma'}_{\!c\,j\sigma}({\bf p'},{\bf p})= 
\sqrt{\frac{M^2}{\ee\ee'}}\,\frac{\sqrt{1-v^2}}{\ee+\ee'-(p_z+p'_z)v}\,
\left[\vd{\sigma'}{p'}V(-{\bf p}-{\bf p'})\u{\sigma}{p}\right].
\la{Wc2}\eeq
Its explicit form in the baryon rest frame has been given in Ref.~\cite{DMinn}.
In the infinite momentum frame (IMF) one has to take the limit $v\to 1$. 
The momentum of the baryon with mass ${\cal M}$ is 
\beq
P=\frac{{\cal M}v}{\sqrt{1-v^2}},\qquad {\rm hence}\;\; 
v=\frac{P}{\sqrt{P^2+{\cal M}^2}}\simeq 1-\frac{{\cal M}^2}{2P^2}.
\la{kin1}\eeq
The quark and the antiquark of the $Q\bar Q$ pair have the 4-momenta
\beq
p_\mu=\left(zP+\frac{p_\perp^2+M^2}{2zP},{\bf p}_\perp, zP\right),
\qquad
p'_\mu=\left(z'P+\frac{p^{'2}_\perp+M^2}{2z'P},{\bf p'}_\perp, z'P\right),
\la{kin2}\eeq
hence the energy denominator is
\beq
\frac{\sqrt{1\!-\!v^2}}{\ee\!+\!\ee'\!-\!(p_z\!+\!p'_z)v}=\frac{{\cal M}}{P}\,
\frac{2zz'P}{Z},\qquad
Z\equiv {\cal M}^2zz'(z+z')+z(p^{'2}_\perp+M^2)+z'(p^{2}_\perp+M^2).
\la{Z}\eeq 
In the infinite momentum frame it is convenient to rescale the annihilation-creation
operators, $a_\sigma^{\rm IMF}(z,{\bf p}_\perp)=\sqrt{P/2\pi}\,\a{\sigma}{p}$
and similarly for $a^\d,b,b^\d$, where the subscript $\sigma=1,2$ refers now
to the $\pm$ helicity states. The new operators satisfy the anticommutation
relations
\beq 
\{a^{\alpha_1f_1\sigma_1}(z_1,{\bf p}_{1\perp}),a^\d_{\alpha_2f_2\sigma_2}(z_2,{\bf p}_{2\perp})\}=
\delta^{\alpha_1}_{\alpha_2}\delta^{f_1}_{f_2}\delta^{\sigma_1}_{\sigma_2}\,
\delta(z_1-z_2)\,(2\pi)^2\delta^{(2)}({\bf p}_{1\perp}-{\bf p}_{2\perp})
\la{antcIMF}\eeq
and similarly for the new $b,b^\d$. The use of the rescaled operators requests 
rescaling $W^{j'\sigma'}_{\!c\,j\sigma}$ by a factor of $P/(2\pi)$. Taking the $v\to 1$
limit in the bispinors \ur{uv} one gets finally~\cite{PP-IMF} \cite{footnote2} 
\beqa\nn
W^{j'\sigma'}_{\!c\,j\sigma}(z,p_\perp;z',p'_\perp)&=&
\frac{M{\cal M}}{2\pi Z}\,\left\{\Sigma^{j'}_j({\bf q})\left[M(z'-z)\sigma_3
+\scp{{\bf Q}_\perp}{\sigma_\perp}\right]^{\sigma'}_\sigma\right.\\
\la{Wc3}
&+&\left.i\,\Pi^{j'}_j({\bf q})\left[-M(z'+z){\bf 1}
-i\ee_{\alpha\beta}Q_{\perp\alpha}\sigma_{\perp\beta}\right]^{\sigma'}_\sigma\right\},\\
\nn
{\bf q}=\left(({\bf p}+{\bf p'})_\perp, (z+z'){\cal M}\right),&&
Q_{\perp\alpha}=zp'_{\perp\alpha}-z'p_{\perp\alpha},\qquad \alpha,\beta=1,2.
\eeqa
The non-primed indices refer here to the quark and the primed ones to the antiquark.
Note that we have written here the conjugate pair wave function. To obtain the pair
wave function actually used in \eq{Psi}, one has to take the hermitian conjugate of $W_{\!c}$, 
namely replace $(j\leftrightarrow j'),\,(\sigma\leftrightarrow \sigma')$ and change 
the sign of the last $i\ee_{\alpha\beta}$ term in \eq{Wc3}:
\beqa\nn
W^{j\sigma}_{j'\sigma'}(z,p_\perp;z',p'_\perp)&=&
\frac{M{\cal M}}{2\pi Z}\,\left\{\Sigma^{j}_{j'}({\bf q})\left[M(z'-z)\sigma_3
+\scp{{\bf Q}_\perp}{\sigma_\perp}\right]^{\sigma}_{\sigma'}\right.\\
&+&\left.i\,\Pi^{j}_{j'}({\bf q})\left[-M(z'+z){\bf 1}
+i\ee_{\alpha\beta}Q_{\perp\alpha}\sigma_{\perp\beta}\right]^{\sigma}_{\sigma'}\right\}
\la{W3}\eeqa
where, again, the primed indices refer to the antiquark. 
   
\Eq{W3} gives the wave function of the additional $Q\bar Q$ pairs in a baryon
in the infinite momentum frame. The corresponding expression for the rest frame is
presented in Ref.~\cite{DMinn}. \Eq{Wc3} is the conjugate wave function needed to
evaluate matrix elements for baryon observables. The indices $j,j'=1,2$ are the isospin
indices (to be rotated by the $SU(3)$ flavor matrices $R$ in \eq{Psi}) and
$\sigma,\sigma'=1,2$ are the quark and antiquark helicity states. The annihilation-creation 
operators in \eq{Psi} are now understood to be normalized by the condition \ur{antcIMF},
and the integrals over momenta there are understood as $\int\!dz\!\int d^2p_\perp/(2\pi)^2$. 

\subsection{Discrete-level wave function}

As seen from \eq{val2}, the discrete-level wave function $F^{j\sigma}({\bf p})
=F^{j\sigma}_{\rm lev}({\bf p})+F^{j\sigma}_{\rm sea}({\bf p})$ consists
of two pieces: one is directly the wave function of the valence level, the other is
related to the change of the number of quarks at the discrete level as due to the presence of
the Dirac sea; it is a relativistic effect and can be ignored in the non-relativistic limit,
together with the lower $L\!=\!1$ component $j(r)$ of the level wave function. Indeed,
in the baryon rest frame the evaluation of the first term in \eq{val2} gives
\beq
F^{j\sigma}_{\rm lev}({\bf p})=\ee^{j\sigma}
\left(\sqrt{\frac{E_{\rm lev}+M}{2E_{\rm lev}}}h(p)+\sqrt{\frac{E_{\rm lev}-M}{2E_{\rm lev}}}j(p)\right),
\la{Flev_rest}\eeq
where $h,j(p)$ are the Fourier transforms of the valence wave functions \ur{level}:
\beqa\la{Fh}
h(p)\!&\!=\!&\!\int\!d^3x\,e^{-i\scp{p}{x}}\,h(r)=4\pi\!\int\!dr\,r^2\frac{\sin\,pr}{pr}\,h(r),\\
\la{Fj}
j^a(p)\!\!&\!=\!&\!\!\!\!\int\!d^3x\,e^{-i\scp{p}{x}}\,(-in^a)j(r)=\frac{p^a}{|p|}j(p),\quad
j(p)=\frac{4\pi}{p^2}\!\int\!dr\,(pr\,\cos\,p r-\sin\,p r)\,j(r).
\eeqa
One sees that the second term in \eq{Flev_rest} is double-suppressed in the non-relativistic
limit $E_{\rm lev}\approx M$: first, owing to the kinematical factor, second, since in this 
limit the $L\!=\!1$ wave $j(r)$ is much less than the $L\!=\!0$ wave $h(r)$. 

In the infinite momentum frame the evaluation of the bispinors $\bar u,\bar v$ from \eq{uv}
produces~\cite{PP-IMF} \cite{footnote3}   
\beq
F^{j\sigma}_{\rm lev}(z,p_\perp)=\sqrt{\frac{{\cal M}}{2\pi}}
\left[\ee^{j\sigma}h(p)+\left(p_z{\bf 1}+i\ee_{\alpha\beta}p_{\perp\alpha}
\sigma_{\perp\beta}\right)^{\sigma}_{\sigma'}\ee^{j\sigma'}\frac{j(p)}{|p|}
\right]_{p_z=z{\cal M}}
\la{Flev_IMF}\eeq

Similarly, the evaluation of the ``sea'' part of the discrete-level wave function
gives
\beq
F^{j\sigma}_{\rm sea}(z,p_\perp)=-\sqrt{\frac{{\cal M}}{2\pi}}
\!\int\!dz'\frac{d^2p'_\perp}{(2\pi)^2}\,W^{j\sigma}_{j'\sigma'}(p,p')\,\ee^{j'\sigma''}\,
\left[(\sigma_3)^{\sigma'}_{\sigma''}h(p')
-(\sigma\cdot{\bf p'})^{\sigma'}_{\sigma''}\frac{j(p')}{|p'|}\right]_{p_z=z{\cal M}}
\la{Fsea_IMF}\eeq
where the pair wave function \ur{W3} has to be used. The conjugate functions are hermitian
conjugate. 

We have thus fully determined all quantities entering the master \eq{Psi} 
for the 3,5,7... Fock components of baryons' wave functions.

\section{Examples of baryon wave functions in the non-relativistic limit}

If the coherent exponent with $Q\bar Q$ pairs is ignored, one gets from the general \Eq{Psi}
the 3-quark Fock component of the octet and decuplet baryons. It depends on the quark ``coordinates'':
the position in space (${\bf r}$), the color ($\alpha$), the flavor ($f$) and the spin ($\sigma$), 
and also on the baryon spin projection $k$. For example, the neutron $3Q$ wave function turns out to be
\beqa\nn
&&\left(|n\!>_k\right)^{f_1f_2f_3,\sigma_1\sigma_2\sigma_3}({\bf r_1,r_2,r_3})
=
\ee^{f_1f_2}\,\ee^{\sigma_1\sigma_2}\,\delta^{f_3}_2\,\delta^{\sigma_3}_k\,
h(r_1)h(r_2)h(r_3)\\
&&+\,{\rm permutations\;of\;1,2,3},
\la{n1}\eeqa
times the antisymmetric $\ee^{\alpha_1\alpha_2\alpha_3}$ in color. It is better known in the form
\beqa\nn
|n\!\uparrow>&\!=\!&2\,d\!\uparrow\!(r_1)d\!\uparrow\!(r_2)
u\!\downarrow\!(r_3)\!-\!d\!\uparrow\!(r_1)u\!\uparrow\!(r_2)
d\!\downarrow\!(r_3)\!-\!u\!\uparrow\!(r_1)d\!\downarrow\!(r_2)
d\!\uparrow\!(r_3)\\
&+& {\rm permutations\;of\;} r_1,r_2,r_3,
\la{n2}\eeqa
which is the well-known non-relativistic $SU(6)$ wave function of the nucleon!
In Ref.~\cite{PP-IMF} the corresponding $SU(6)$ function in the infinite momentum frame
was obtained. 

Similarly, the $3Q$ component of the $\Delta^0$ baryon with spin projection 1/2, whose
wave function may be compared with that of the neutron, is
\beqa
\nn
&&|\Delta^0\!\uparrow>^{f_1f_2f_3,\sigma_1\sigma_2\sigma_3}({\bf
r_1,r_2,r_3})= 
\left(\delta^{f_1}_1\delta^{f_2}_2\delta^{f_3}_2
+\delta^{f_1}_2\delta^{f_2}_1\delta^{f_3}_2
+\delta^{f_1}_2\delta^{f_2}_2\delta^{f_3}_1\right)\\
\la{Delta01}
&&\cdot\left(\delta^{\sigma_1}_1\delta^{\sigma_2}_1\delta^{\sigma_3}_2
+\delta^{\sigma_1}_1\delta^{\sigma_2}_2\delta^{\sigma_3}_1
+\delta^{\sigma_1}_2\delta^{\sigma_2}_1\delta^{\sigma_3}_1\right)
h(r_1)h(r_2)h(r_3)
\eeqa
which can be presented also as a familiar $SU(6)$ wave function
\beqa
\nn
|\Delta^0\!\uparrow>&=&
 u\!\uparrow\!(r_1)\,d\!\uparrow\!(r_2)\,d\!\downarrow\!(r_3)+
 d\!\downarrow\!(r_1)\,u\!\uparrow\!(r_2)\,d\!\uparrow\!(r_3)+
 d\!\uparrow\!(r_1)\,d\!\uparrow\!(r_2)\,u\!\downarrow\!(r_3)\\
 &+&{\rm permutations\;of\;} r_1,r_2,r_3.
 \la{Delta02}\eeqa
There are, of course, relativistic corrections to these $SU(6)$-symmetric
formulae, arising from i) exact treatment of the discrete level, \eqs{Flev_IMF}{Fsea_IMF},
and ii) additional $Q\bar Q$ pairs described by \eq{W3}. Both effects are not small.  

The $5Q$ component of a baryon is obtained when one expands the coherent
exponent to the linear order and then projects it into the concrete baryon in question.
In general, the $5Q$ wave functions look rather complicated as they depend on 
five quark ``coordinates'', including their coordinates proper, spin, flavor and color. 
We do not write explicitly the color degrees of freedom but always imply that the $(1,2,3)$
quarks of the level are antisymmetric in color while the quark-antiquark pair $(4,5)$ 
is a color singlet, as it follows from \eq{Psi}. For example, the $5Q$ component of 
the neutron has the wave function
\beqa
\nn
&&\left(|n\!>_k\right)^{f_1f_2f_3f_4,\sigma_1\sigma_2\sigma_3\sigma_4}_{f_5,\sigma_5}
({\bf r_1,r_2,r_3,r_4,r_5})=
F^{j_1\sigma_1}({\bf r_1})F^{j_2\sigma_2}({\bf r_2})F^{j_3\sigma_3}({\bf r_3})
W^{j_4\sigma_4}_{j_5\sigma_5}({\bf r_4,r_5})\\
\nn
&\cdot &
\!\left\{\ee^{f_1f_2}\epsilon_{j_1j_2}\left[\delta^{f_3}_2\delta^{f_4}_{f_5}
\left(4\delta^{j_5}_{j_4}\delta^{k^\prime}_{j_3}
-\delta^{j_5}_{j_3}\delta^{k^\prime}_{j_4}\right)+
\delta^{f_4}_2\delta^{f_3}_{f_5}\left(4\delta^{j_5}_{j_3}\delta^{k^\prime}_{j_4}
-\delta^{j_5}_{j_4}\delta^{k^\prime}_{j_3}\right)\right]\right.\\
\nn
&+&
\!\!\left.\ee^{f_1f_4}\epsilon_{j_1j_4}\left[\delta^{f_2}_2\delta^{f_3}_{f_5}
\left(4\delta^{j_5}_{j_3}\delta^{k^\prime}_{j_2}\!
-\!\delta^{j_5}_{j_2}\delta^{k^\prime}_{j_3}\right)+
\delta^{f_3}_2\delta^{f_2}_{f_5}
\left(4\delta^{j_5}_{j_2}\delta^{k^\prime}_{j_3}\!
-\!\delta^{j_5}_{j_3}\delta^{k^\prime}_{j_2}\right)\right]\right\}\epsilon_{k^\prime k}
\\
&+&\!{\rm permutations\; of}\; (1,2,3)
\la{n5}\eeqa
where the pair wave function $W$ in the rest frame can be found in Ref.~\cite{DMinn}.
Indices 1-3 refer to quarks at the discrete level, 4 refers to
the quark in the additional pair, and 5 refers to the antiquark in
the pair. Terms of the type of $\delta^{f_3}_{f_5}$ mean the flavor-symmetric
combination $s\bar s+u\bar u+d\bar d$. We have not invented how to present 
it in a more compact form. 

Turning to the exotic baryons from the $\left(\at,\half^+\right)$, 
projecting the three quarks from the discreet level on the $\Theta$ rotational 
function \ur{Theta} gives an identical zero (see \eq{no3quarks}), in accordance with 
the fact that the $\Theta$ cannot be made of 3 quarks. The non-zero projection is achieved
when one expands the coherent exponent at least to the linear order. One gets then 
from \eq{RTheta} the $5Q$ component of the $\Theta$ wave function : 
\beqa\nn
&&|\Theta^+_k\!>^{f_1f_2f_3f_4,\sigma_1\sigma_2\sigma_3\sigma_4}_{f_5,\sigma_5}({\bf r_1\ldots r_5})
=
\ee^{f_1f_2}\ee^{f_3f_4}\delta^3_{f_5}\,\ee^{\sigma_1\sigma_2}\\
&&\cdot\, h(r_1)h(r_2)h(r_3)\,
W^{\sigma_3\sigma_4}_{k\,\sigma_5}({\bf r_4,r_5})+{\rm permutations\;of\;1,2,3}
\la{Theta1}\eeqa
The color structure of the antidecuplet wave function is
$\ee^{\alpha_1\alpha_2\alpha_3}\delta^{\alpha_4}_{\alpha_5}$.
Indices 1 to 4 refer to quarks and index 5 refers to the antiquark,
in this case $\bar s$ owing to $\delta^3_{f_5}$. The quark
flavor indices are $f_{1\!-\!4}=1,2=u,d$. Naturally, we have obtained
$\Theta^+=uudd\bar s$. Since in the CQSM the functions $h(r_{1,2,3})$ and 
$W({\bf r_4},{\bf r_5})$ are known, \eq{Theta1} gives the complete
color, flavor, spin and space 5-quark wave function of the $\Theta^+$ in its rest
frame. The structure $\ee^{f_1f_2}\ee^{\sigma_1\sigma_2}$ clearly shows that there
is a pair of $ud$ quarks in the spin and isospin zero combination, exactly
as in the nucleon, \eq{n1}. However, it does not mean that there are prominent 
scalar isoscalar diquarks either in the nucleon or in the $\Theta$: 
that would require their spatial correlation which, as we see, is absent 
in the mean field approximation. 

The $Q\bar Q$ pair wave function $W$ is a combination of four partial waves with
different permutation symmetries. The partial waves depend separately on the
coordinates ${\bf r_{4,5}}$ measured from the baryon center of mass. 
More explicit formulae can be found in Ref.~\cite{DMinn}. 
It would be interesting to compare \eq{Theta1} with the wave functions 
obtained in non-relativistic dynamical models or discussed in that framework~\cite{nrm}. 

Unfortunately, the meaning of the baryon wave function in the rest frame 
is unclear since it has fundamental flaws, especially in the world with 
the spontaneous chiral symmetry breaking and light pions, as we have had several 
chances to explain~\cite{DP-mixing,D04}. In particular, a rotation of the 
chiral phase is a zero-energy vacuum rearrangement in the chiral limit, however it
can be decomposed into a large number of $Q\bar Q$ pairs. To find the true
number of $Q\bar Q$ pairs in a hadron, one has to separate its proper structure 
from that of the vacuum. To this end one is forced to consider hadrons in the 
infinite momentum frame where the vacuum $Q\bar Q$ pairs with an arbitrary
high momenta are nevertheless separated from those belonging to a hadron and
thus having an infinite momentum. Therefore, in the IMF (and only there)
the number of $Q\bar Q$ pairs in a hadron has a precise mathematical meaning,
and the Fock states are well defined.

\section{Three quarks: normalization, vector and axial charges}

The normalization of a baryon wave function in the second-quantization representation
\ur{Psi} is found from 
\beq
{\cal N}_B=\frac{1}{2}\delta^k_l\,<\!\Psi^{\d\,B\,l}\Psi^B_k\!>.
\la{N1}\eeq 
The annihilation operators in $\Psi^{\d\,B\,l}$ must be dragged to the right where 
they ultimately nullify the vacuum state $|0\!\!>$ and the creation operators 
from $\Psi^B_k$ should be dragged to the left where they ultimately nullify 
the vacuum state $<\!\!0|$. The result is non-zero owing to the anticommutation 
relations \ur{antcIMF} or the ``contractions'' of the operators. 

For the $3Q$ Fock component of a baryon, there are $3!$ possible (and equivalent) 
contractions, and the ensuing contraction in color indices gives another factor
of $3!=\ee^{\alpha_1\alpha_2\alpha_3}\ee_{\alpha_1\alpha_2\alpha_3}$. 
Flavor projecting to a baryon with specific quantum numbers gives a tensor
\beq
T^{f_1f_2f_3}_{j_1j_2j_3,k}=\int\!dR\,B^*_k(R)\,R^{f_1}_{j_1}R^{f_2}_{j_2}R^{f_3}_{j_3}
\la{T3}\eeq 
and a hermitian conjugate for the conjugate wave function. Hence the normalization
of the $3Q$ component is
\beqa\nn
&&{\cal N}^{(3)}=\frac{(6\!\cdot\!6)\!}{2}\,\delta^k_l\,
T^{f_1f_2f_3}_{j_1j_2j_3,k}\,T^{l_1l_2l_3,l}_{f_1f_2f_3}
\int\!dz_{1,2,3}\!\int\frac{d^2p_{1,2,3}}{(2\pi)^6}\,\delta(z_1+z_2+z_3-1)\\
&&\cdot (2\pi)^2 \delta((p_1\!+\!p_2\!+\!p_3)_\perp)\,
F^{j_1\sigma_1}(p_1)F^{j_2\sigma_2}(p_2)F^{j_3\sigma_3}(p_3)
\,F^\d_{l_1\sigma_1}(p_1)F^\d_{l_2\sigma_2}(p_2)F^\d_{l_3\sigma_3}(p_3)
\la{N31}\eeqa
where $F^{j\sigma}(z,p_{\perp})$ are the level wave functions \urs{Flev_IMF}{Fsea_IMF}. 
In the non-relativistic limit $F^{j\sigma}(p)F^\d_{l\sigma}(p)=\delta^j_l\,h^2(p)$. Therefore
in this simple case the normalization is the full contraction of the two $T$ tensors, times
an integral over momenta which can be performed numerically once the level wave function
$h(p)$ is known. Since the normalization of this function is arbitrary one can always choose
it such that ${\cal N}^{(3)}=1$ for all baryons possessing a $3Q$ component. 

A typical physical observable is a matrix element of some operator (which should
be written down in terms of the quark annihilation-creation operators $a,b,a^\d,b^\d$)
sandwiched between initial and final baryon wave functions \ur{Psi}. We shall consider as
examples the operators of the vector and axial charges which can be written through the 
annihilation-creation operators as 
\beqa\nn
\left\{\ba{c}Q\\Q_5\ea\right\}\!\!\!&=&\!\!\!\int\!d^3x\,\bar\psi_e\,J^e_h\,
\left\{ \ba{c} \gamma_0 \\ 
\gamma_0\gamma_5 \ea \right\} \psi^h
=\int\!dz\,\frac{d^2p_\perp}{(2\pi)^2}
\left[
a^\d_{e\pi}(z,p_\perp)a^{h\rho}(z,p_\perp)\,J^e_h
\left\{\ba{c}\delta^{\pi}_{\rho}\\(-\sigma_3)^{\pi}_{\rho}\ea\right\}
\right.
\\
&-&\left.
b^{\d\,h\rho}(z,p_\perp)b_{e\pi}(z,p_\perp)\,J^e_h
\left\{\ba{c}\delta^{\pi}_{\rho}\\(-\sigma_3)^{\pi}_{\rho}\ea\right\}
\right] 
\la{vac}\eeqa
where $J^e_h$ is the flavor contents of the charge, and $\pi,\rho=1,2$ are helicity
states. For example, if
we consider the $\rho^+=\bar d u$ current which annihilates $u$ quarks and creates
$d$ quarks and annihilates $\bar d$ antiquarks and creates $\bar u$ ones, the flavor
currents in \eq{vac} are $J^e_h(\rho^+)=\delta^e_2\delta^1_h$. Notice that there are no 
$a^\d b^\d$ or $ab$ terms in the charges. This is a great advantage of the IMF 
where the number of $Q\bar Q$ pairs is not changed by the current. Hence there 
will be only diagonal transitions between Fock components with equal numbers of pairs. 

\begin{figure}[htb]
\begin{minipage}[t]{.45\textwidth}
\includegraphics[width=\textwidth]{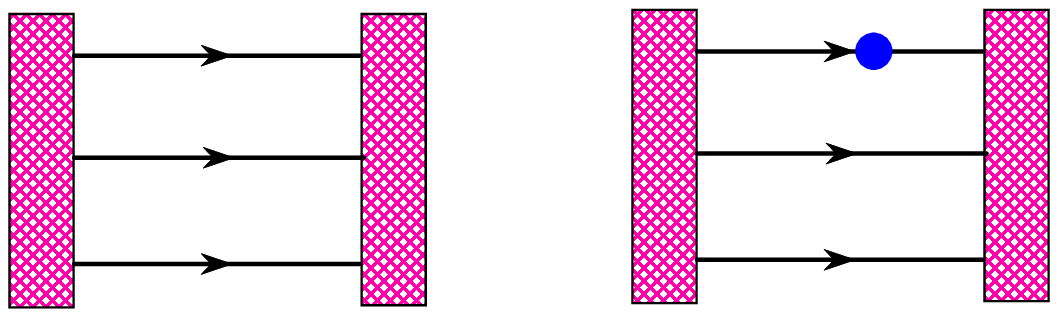}
\caption{Graphs showing the normalization of a 3-quark component of a baryon (left)
and the matrix element of a local operator denoted by a circle (right).}
\label{fig:2}
\end{minipage}
\hfil
\begin{minipage}[t]{.40\textwidth}
\includegraphics[width=\textwidth]{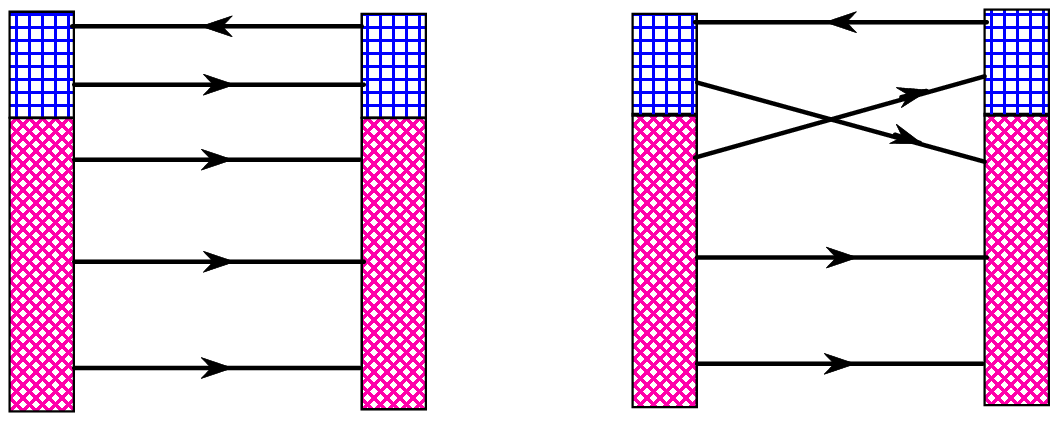}
\caption{Direct (left) and exchange (right) contributions to the normalization 
of the 5-quark component of a baryon. The upper rectangles denote $Q\bar Q$ pairs.}
\label{fig:3}
\end{minipage}
\end{figure}

In the matrix elements between the $3Q$ components the $b^\d b$ part of the current 
is passive as there are no antiquarks. The $a^\d a$ part is a sum over colors. 
As in the normalization, one gets the factor $6\cdot 6$ from all contractions. Let it 
be the third quark whose charge is measured: there is a factor of 3 from three quarks to 
which the charge operator can be applied, see Fig.~3. Denoting for short 
$\int\!(dp_{1-3})$ the integrals over momenta with the conservation $\delta$-functions 
as in \eq{N31} we arrive at the following expression for the matrix element of the vector charge: 
\beqa\nn
V^{(3)}&=&\frac{(6\!\cdot\!6\!\cdot\!3)\!}{2}\,\delta^k_l\,T(1)^{f_1f_2f_3}_{j_1j_2j_3,k}\,
T(2)^{l_1l_2l_3,l}_{f_1f_2g_3}\int\!(dp_{1-3})\\
&&\cdot \left[F^{j_1\sigma_1}(p_1)F^{j_2\sigma_2}(p_2)F^{j_3\sigma_3}(p_3)\right]
\,\left[F^\d_{l_1\sigma_1}(p_1)F^\d_{l_2\sigma_2}(p_2)F^\d_{l_3\tau_3}(p_3)\right]\,
\left[\delta^{\tau_3}_{\sigma_3}J^{g_3}_{f_3}\right].
\la{V31}\eeqa
One can easily check using \eqs{p}{n} that, say, for the $p\to n\rho^+$ transition, the above
vector charge gives exactly the same expression as for the normalization \ur{N31}. Therefore,
the $g_V$ of this transition is unity, as it should be for the conserved vector current. 

For the axial transition, one replaces averaging over baryon spin by $\half(-\sigma_3)^k_l$, 
and the axial charge operator is now $(-\sigma_3)^{\tau_3}_{\sigma_3}$ instead of 
$\delta^{\tau_3}_{\sigma_3}$, see \eq{vac}. All the rest is the same as in \eq{V31}: 
\beqa\nn
A^{(3)}&=&\frac{(6\!\cdot\!6\!\cdot\!3)\!}{2}\,(-\sigma_3)^k_l\,T(1)^{f_1f_2f_3}_{j_1j_2j_3,k}\,
T(2)^{l_1l_2l_3,l}_{f_1f_2g_3}\int\!(dp_{1-3})\\
&&\!\cdot\!\left[F^{j_1\sigma_1}(p_1)F^{j_2\sigma_2}(p_2)F^{j_3\sigma_3}(p_3)\right]
\left[F^\d_{l_1\sigma_1}(p_1)F^\d_{l_2\sigma_2}(p_2)F^\d_{l_3\tau_3}(p_3)\right]\!
\left[(\!-\sigma_3)^{\tau_3}_{\sigma_3}J^{g_3}_{f_3}\right]\!.
\la{A31}\eeqa
The result, however, is now different as the axial charge is not conserved. 
For example, for the $p\to n\pi^+$ transition one gets
the expression identical to that for the normalization but with the factor 5/3. It means that 
we have obtained in the non-relativistic limit for the $3Q$ component of the nucleon
$g_{A}^{(3)}(N)=5/3$. It is the well-known result of the non-relativistic quark model. 
However, it is modified by the relativistic corrections to the valence quark
wave functions \urs{Flev_IMF}{Fsea_IMF} and by the $5Q$ component of the nucleon.

\section{Five quarks: normalization, vector and axial charges}

Already in the normalization of the $5Q$ Fock component of a baryon there are two types
of contributions: direct and exchange ones, see Fig.~4. In the former, one contracts $a^\d$ from the pair
wave function with an $a$ in the conjugate pair, and all the valence operators are contracted
with each other. There are 6 such possibilities, and the contraction in color gives a factor
$3\cdot 6$, all in all 108. In the exchange contributions, one contracts $a^\d$ from the pair
with one of the three $a$'s from the valence level. Further on, $a$ from the conjugate pair
is contracted with one of the three $a^\d$'s from the valence level. There are 18 such 
possibilities but the contraction in color gives now only a factor of 6. Therefore for the exchange
contractions we also get a factor of 108 but with an overall negative sign as one has to
anticommute fermion operators to get the exchange terms. As a result we obtain the
following general expression for the normalization of the $5Q$ Fock component:
\beqa
\nn
&&
{\cal N}^{(5)}=\frac{108}{2}\int\!(dp_{1-5})\,\delta^k_l\,T^{f_1f_2f_3f_4,j_5}_{j_1j_2j_3j_4,f_5,k}\,
T^{l_1l_2l_3l_4,f_5,l}_{f_1f_2g_3g_4,l_5}\\
\nn
&\cdot &F^{j_1\sigma_1}(p_1)F^{j_2\sigma_2}(p_2)F^{j_3\sigma_3}(p_3)\,
W^{j_4\sigma_4}_{j_5\sigma_5}(p_4,p_5)\,F^\dagger_{l_1\sigma_1}(p_1)F^\dagger_{l_2\sigma_2}(p_2)\\
\la{N51}
&\cdot &\left[F^\dagger_{l_3\sigma_3}(p_3)W^{l_5\sigma_5}_{\!c\,l_4\sigma_4}(p_4,p_5)\,
\delta^{g_3}_{f_3}\delta^{g_4}_{f_4}
-F^\dagger_{l_3\sigma_4}(p_4)W^{l_5\sigma_5}_{\!c\,l_4\sigma_3}(p_3,p_5)
\delta_{f_4}^{g_3}\delta_{f_3}^{g_4}\right].
\eeqa
The flavor tensor here is the group integral projecting the $5Q$ state onto a particular baryon:
\beq
T^{f_1f_2f_3f_4,j_5}_{j_1j_2j_3,f_5,k}=\int\!dR\,B^*_k(R)\,R^{f_1}_{j_1}R^{f_2}_{j_2}R^{f_3}_{j_3}
R^{f_4}_{j_4}R^{\d\,j_5}_{f_5}.
\la{T5}\eeq 

\begin{figure}[htb]
\begin{minipage}[t]{.40\textwidth}
\includegraphics[width=\textwidth]{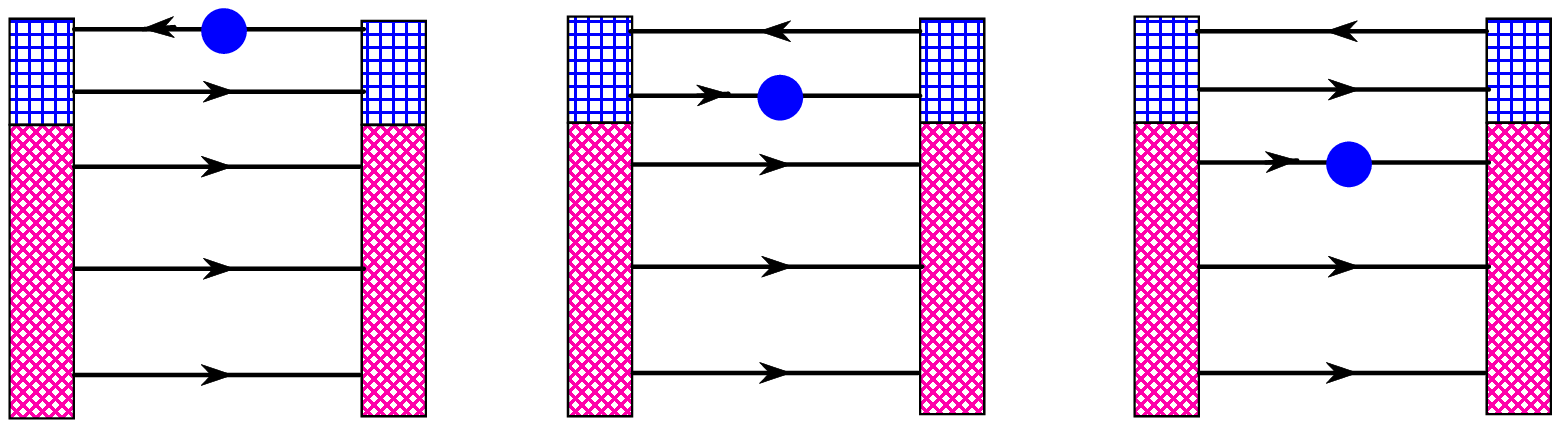}
\caption{Direct contributions to the matrix element of an operator, 
in the 5-quark component of a baryon. The operator is applied to the antiquark (left), 
to the quark from the pair (middle) and to the quark from the valence level (right).}
\label{fig:2}
\end{minipage}
\hfil
\begin{minipage}[t]{.50\textwidth}
\includegraphics[width=\textwidth]{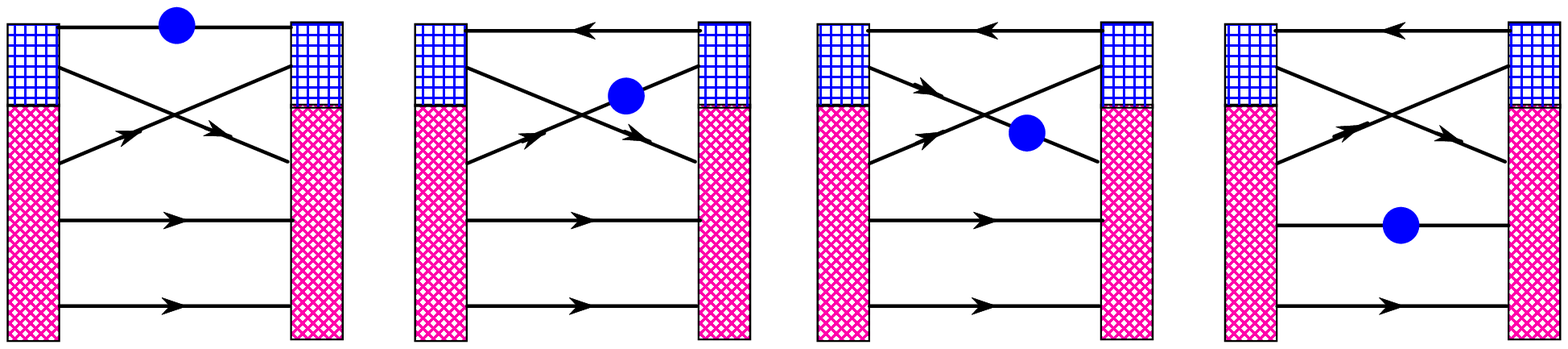}
\caption{Four types of exchange contributions to the matrix element in the 5-quark
component of a baryon.}
\label{fig:3}
\end{minipage}
\end{figure}

The ratio of the normalizations ${\cal N}^{(5)}/{\cal N}^{(3)}$ gives the probability
to find a $5Q$ component in a mainly $3Q$ baryon. It depends on the mean field inside
a baryon through the pair wave function $W$ (and is quadratic in the mean field), 
and on the particular baryon through its spin-flavor contents $T$.  

For the vector and axial transitions there are three basic contributions: one when the charge
of the antiquark is measured, the second when the charge operator acts on the quark
from the pair, and the third when it acts on one of the three valence quarks. 
These three types are further divided into the direct and exchange contributions (Figs.~5,6). 
We write below only the direct contributions.

The vector transition:
\beqa
\nn
&&
V^{(5){\rm direct}}=\frac{108}{2}\int\!(dp_{1-5})\,\delta^k_l\,
T(1)^{f_1f_2f_3f_4,j_5}_{j_1j_2j_3j_4,f_5,k}\,
T(2)^{l_1l_2l_3l_4,g_5,l}_{f_1f_2g_3g_4,l_5}\\
\nn
&\cdot &F^{j_1\sigma_1}(p_1)F^{j_2\sigma_2}(p_2)F^{j_3\sigma_3}(p_3)\,
W^{j_4\sigma_4}_{j_5\sigma_5}(p_4,p_5)\,F^\dagger_{l_1\sigma_1}(p_1)
F^\dagger_{l_2\sigma_2}(p_2)F^\dagger_{l_3\tau_3}(p_3)\,
W^{l_5\tau_5}_{\!c\,l_4\tau_4}(p_4,p_5)\\
\la{V51}
&\cdot &\left[-\delta_{f_3}^{g_3}\delta_{f_4}^{g_4}J_{g_5}^{f_5}\delta^{\tau_3}_{\sigma_3}
\delta^{\tau_4}_{\sigma_4}\delta^{\sigma_5}_{\tau_5}
+\delta_{f_3}^{g_3}J_{f_4}^{g_4}\delta_{g_5}^{f_5}\delta^{\tau_3}_{\sigma_3}
\delta^{\tau_4}_{\sigma_4}\delta^{\sigma_5}_{\tau_5}+
3J_{f_3}^{g_3}\delta_{f_4}^{g_4}\delta_{g_5}^{f_5}\delta^{\tau_3}_{\sigma_3}
\delta^{\tau_4}_{\sigma_4}\delta^{\sigma_5}_{\tau_5}\right].
\eeqa

The axial transition:
\beqa
\nn
&&
A^{(5){\rm direct}}=\frac{108}{2}\int\!(dp_{1-5})\,(-\sigma_3)^k_l\,
T(1)^{f_1f_2f_3f_4,j_5}_{j_1j_2j_3j_4,f_5,k}\,
T(2)^{l_1l_2l_3l_4,g_5,l}_{f_1f_2g_3g_4,l_5}\\
\nn
&\cdot &F^{j_1\sigma_1}(p_1)F^{j_2\sigma_2}(p_2)F^{j_3\sigma_3}(p_3)\,
W^{j_4\sigma_4}_{j_5\sigma_5}(p_4,p_5)\,F^\dagger_{l_1\sigma_1}(p_1)
F^\dagger_{l_2\sigma_2}(p_2)F^\dagger_{l_3\tau_3}(p_3)\,
W^{l_5\tau_5}_{\!c\,l_4\tau_4}(p_4,p_5)\\
\la{A51}
&\cdot &\left[\delta_{f_3}^{g_3}\delta_{f_4}^{g_4}J_{g_5}^{f_5}\delta^{\tau_3}_{\sigma_3}
\delta^{\tau_4}_{\sigma_4}(\sigma_3)^{\sigma_5}_{\tau_5}
-\delta_{f_3}^{g_3}J_{f_4}^{g_4}\delta_{g_5}^{f_5}\delta^{\tau_3}_{\sigma_3}
(\sigma_3)^{\tau_4}_{\sigma_4}\delta^{\sigma_5}_{\tau_5}
-3J_{f_3}^{g_3}\delta_{f_4}^{g_4}\delta_{g_5}^{f_5}(\sigma_3)^{\tau_3}_{\sigma_3}
\delta^{\tau_4}_{\sigma_4}\delta^{\sigma_5}_{\tau_5}\right],
\eeqa
where $J^e_h$ is the flavor content of the current defined in the previous section.
 
Applications of these general formulae to physically interesting cases, for example
for the calculation of the $\Theta^+$ width, will be presented in a subsequent publication.

\section{Conclusions}

We have presented a technique allowing to write down explicitly the quark wave functions 
of the octet, decuplet and antidecuplet baryons, in the mean field approximation. Having
patience (and space) one can write down the 19-quark component of the proton or the 7-quark
component of the exotic $\Xi^{-\,-}$. This technique is mathematically equivalent to
the ``valence quarks plus Dirac continuum" method exploited previously, but brings
the mean field approach even closer to the language of the quark wave functions used by many 
people. We have shown that the standard $SU(6)$ wave functions are easily reproduced
for the octet and decuplet baryons, if one assumes the non-relativistic limit. However,
we have given explicit formulae for the relativistic corrections to the $3Q$ wave function,
and also explicitly and for the first time, the $5Q$ wave function of the nucleon and
of the exotic $\Theta^+$.    
  
There seems to be a broad field of applications. One has been already started in
Ref.~\cite{PP-IMF} and involves distribution amplitudes, exclusive processes, 
parton distributions and the like. The other, probably even more broad, is for low energies. 
One can compute any kind of transition amplitudes between baryons, including effects of $SU(3)$
symmetry violation, mixing between multiplets and the widths. \\ 

\vskip 0.4true cm  
{\bf Acknowledgement}\\

D.D.'s work has been supported in part by the DOE under contract DE-AC05-84ER40150.


\begin{thebibliography}{100}
\bibitem{footnote2}
The present pair wave function is that of Ref.~\cite{PP-IMF} multiplied by a constant matrix $\sigma_{\,3}$.
\bibitem{footnote3}
This expression differs from that of Ref.~\cite{PP-IMF}
by a multiplication by the matrix $\sigma_{\,3}$. Also, $h(p),j(p)$ used here are
those used in Ref.~\cite{PP-IMF} divided by $p^2$. This comment refers also to \eq{Fsea_IMF}.
\end{thebibliography}
\end{document}